\documentclass{iopart}
\usepackage{graphicx}
\begin{document}
\title{Casimir energy in spherical cavities}
\author{W. Luis Moch\' an$^1$ and Carlos Villarreal-Luj\'an$^2$} 
\address{$^1$Instituto de Ciencias F\'{\i}sicas, Universidad Nacional
  Aut\'onoma de M\'exico, Apartado Postal 48-3, 62251 Cuernavaca,
  Morelos, M\'exico.}
\address{$^2$Instituto de F\'{\i}sica, Universidad Nacional 
Aut\'onoma
de M\'exico, Apartado Postal 20-364, 01000 Distrito Federal, M\'exico.}
\eads{\mailto{WLM <mochan@fis.unam.mx>}, \mailto{CV <carlos@fisica.unam.mx>}} 
%Article id. A/263343/SPE ??
\begin{abstract}

We calculate the Casimir energy  at spherical cavities
within a host made up of an arbitrary material described by a possibly
dispersive and lossy dielectric response. To that end, we add to the
coherent optical response a contribution that takes account of the
incoherent radiation emitted by the host in order to guarantee the
detailed balance required to keep the system at thermodynamic
equilibrium in the presence of dissipation. The resulting boundary
conditions allow a conventional quantum mechanical treatment of the
radiation within the cavity from which we obtain the contribution of
the cavity walls to the density of states, and from it, the
thermodynamic properties of the system. The contribution of the cavity
to the energy diverges as it incorporates the interaction energy
between neighbor atoms in a continuum description. The change in the
energy of an atom situated at the center of the cavity due to its
interaction with the fluctuating cavity field is however finite. We evaluate
the latter for a simple case.

\end{abstract}

\section{Introduction}

Motivated by his finding that zero-point fluctuations may induce an
attractive force between parallel conducting plates \cite{casimir48},
Casimir proposed in 1956 that the zero-point force could be the
Poincar\'e stress involved a semiclassical model of the electron
\cite{casimir56}. In that model the electron was considered as a
spherical charge distribution stabilized by vacuum fluctuations.
However, T. H. Boyer \cite{boyer68} showed in 1968 that the stress for
a spherical conducting shell of radius $a$ is indeed repulsive, since
the Casimir energy turns out to be positive: $E=.09235/2a$. Subsequent
calculations based on the Green's function method \cite{milton78}, or
a multiple scattering formalism \cite{balian78} confirmed Boyer's
calculation.  The more general problem of the Casimir effect of a
dielectric ball was first considered by Milton \cite{milton80} in
absence of dispersion, and later on by Candelas \cite{candelas82}.
Candelas argued that in presence of boundaries vacuum energies depend
on a cutoff on transverse momenta, independently of the dielectric
properties of the media.  Therefore, Boyer's result should be
corrected by terms associated to surface and curvature tensions.  The
Casimir forces for a dilute dielectric and diamagnetic sphere was
studied by Brevik and Kolventvedt \cite{brevikol}, and more recently
by Klich \cite{klich00}, while the role of dispersion in this problem
was discussed in Refs.\cite{brevik88,brevik89,brevik94}.  In this
case, the Casimir stress may be $attractive$, but very sensitive to
the specific values of the parameters characterizing the electric and
magnetic response of the materials.  An excellent review of different
approaches to the Casimir effect of spherical regions, together with
applications to QCD bag models, higher dimensional spaces, or
sonoluminescence can be found in Ref. \cite{milton}.

In a series of papers \cite{esquivel05,mochan05,njp} an
expression for the Casimir force within planar cavities was derived
without making 
particular models or assumptions about the nature of the walls. By
considering that the system is in thermodynamic equilibrium, we
obtained the energy and the stress tensor in a closed ancillary system
that has the same optical response as the original system. This
approach consistently incorporates evanescent fields and allows a
fully quantum-mechanical treatment of the electromagnetic degrees of
freedom. Unlike the calculations presented in
\cite{esquivel05,mochan05,njp} in Ref.\cite{rmf} the
fictitious system was eliminated, keeping only its essential property:
that detailed balance
should hold in thermodynamic equilibrium: for each photon that is not
coherently reflected at the 
cavity walls and is therefore either absorbed or transmitted beyond the
system, an identical photon has to be incoherently injected back into
the cavity with no phase relation with the lost photon. 
In this paper we show that that approach may be
straightforwardly generalized to study the Casimir effect of spherical
cavities with arbitrary dielectric properties.  We then apply the
formalism to calculate the energy shift for a polarizable atom placed at
the center of a cavity.
  
\section{Theory}
\begin{figure}
\centering{\includegraphics{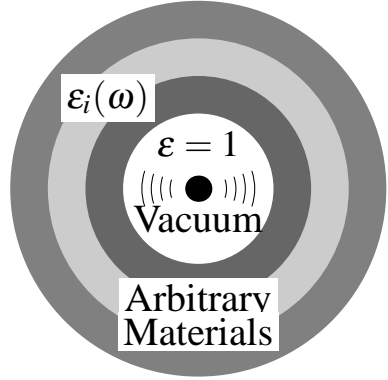}}
\caption{\label{cavity}Spherical cavity within arbitrary system with a
  scatterer at its center.}
\end{figure}

Consider a system made up of an empty spherical cavity of
radius $R$ carved out of an arbitrary material and with a scatterer
situated at its center (Fig. \ref{cavity}).
Within the empty space of the cavity there are outgoing  ($o$) and
ingoing ($i$) electromagnetic waves with transverse electric (TE) and
transverse magnetic (TM) polarization, described by the field
\begin{equation}\label{Flm}
  F_{lm}^{d\zeta}(\vec r) = f_{lm}^{d\zeta}h_l^d(kr)
  Y_{lm}(\hat n), 
\end{equation}
where $d=o,i$ describes the propagation direction,
$\zeta=\mathrm{TE}, \mathrm{TM}$ describes the polarization,
$l=0,1,2,\ldots$ denotes the angular momentum, $m=-l\ldots l$ its
projection along the $z$ axis, $h_l^o\equiv h_l^{(1)}$ and
$h_l^i\equiv h_l^{(2)}$ are  the outgoing and the ingoing spherical Hankel
functions respectively, $k=\omega/c$ is the wavenumber within vacuum,
and we choose the scalar field  
$F_{lm}^{d,\mathrm{TE}}\equiv \vec B_{lm}^{d,\mathrm{TE}}\cdot
\vec r$ to describe the TE electromagnetic field $\vec E_{lm}^{TE}$,
$\vec B_{lm}^{TE}$ and 
$F_{lm}^{d,\mathrm{TM}}\equiv \vec E_{lm}^{d,\mathrm{TM}}\cdot
\vec r$ to describe the TM electromagnetic field $\vec E_{lm}^{TM}$,
$\vec B_{lm}^{TM}$ \cite{jackson}. When the outgoing radiation reaches
the boundary of the cavity it is partially scattered back into the
cavity with an amplitude $s_{b,lm}^\zeta$, and when the ingoing
radiation hits the scatterer at the center it is scattered back
towards the cavity with an amplitude $s_{c,lm}^\zeta$, that is,
\begin{equation}\label{sb}
  f^i=s_b f^o,
\end{equation}
\begin{equation}\label{sc}
  f^o=s_c f^i,
\end{equation}
where we removed the indices $l$, $m$, and $\zeta$ to simplify our
notation. 
From Eqs. (\ref{sb}) and (\ref{sc}) we immediately obtain the {\em
  normal modes} of the system, given by
\begin{equation}\label{normal}
  1-s_b s_c=0.
\end{equation}
Taking into account the frequency ($\omega$) dependence of $s_b$ and
$s_c$ we may solve Eq. (\ref{normal}) to obtain the frequency spectrum
$\omega_n$ for each value of $l,m,\zeta$. However, in the presence of
absorbing materials or even of transparent, leaky materials, $\omega_n$
would turn out to be complex. The 
system would be {\em open} and ordinary quantum mechanics would not be
applicable to the radiation field, i.e.,  $\hbar \omega_n$ would not
be an energy quantum. 

The coefficients $s_b$ and $s_c$ describe the amplitude and the phase of
the radiation that is coherently  scattered back into the cavity.
% and
%they could be measured by introducing a radiation source into the
%cavity, illuminating either its boundary or its center and measuring
%the scattered field. This would be an out of equilibrium measurement.
%
The energy that is not coherently scattered back
into the cavity, described by $1-|s_c|^2$ and $1-|s_b|^2$, is
absorbed or it leaves the
system through its external boundary. Nevertheless, in thermodynamic
equilibrium, an absorbing scatterer or an absorbing enclosure has to
eventually radiate back any radiation that it absorbs. The system
should also admit photons from the vacuum that surrounds it to
replenish those photons that were transmitted away. Detailed balance
must hold and in the average, for each photon with numbers $l,m,\zeta$
that leaves the cavity, an  identical photon, with the same numbers but
having no phase relation with the original photon, has to be injected
back into the cavity. We mimic this incoherent radiation by a coherent
field that is delayed a large time $T_\chi$ ($\chi=b,c$). To avoid 
interference with the coherently scattered field we take the
limit $T_\chi\to\infty$. Thus, the incoherent radiation may be taken into
account by replacing the scattering coefficients $s_b$ and $s_c$ by
{\em total} scattering coefficients
\begin{equation}
  s_\chi\to S_\chi=\frac{s_\chi+a_\chi e^{i\omega T_\chi}} {1 + b_\chi
    e^{i\omega T_\chi}},\quad \chi=b,c.
\end{equation}
Here, $\exp(i\omega T_\chi)$ is the phase acquired during
the large delay $T_\chi$ and is an extremely fast varying function of
the frequency $\omega$, so that for any finite bandwidth
interference effects would disappear. We assume that $a_\chi$ and $b_\chi$ are
relatively slowly varying functions of frequency which are to be
determined. The term $s_\chi$ in the numerator corresponds to the
coherent scattering.  The term $a_\chi e^{i\omega T_\chi}$ corresponds
to re-radiation by the central scatterer ($\chi=c$), re-radiation by the walls of
the cavity ($\chi=b$), or to photons that enter the system from
outside to replenish the cavity losses. Finally, it may happen that a
re-radiated photon fails to reach the cavity on its first
attempt, as it may be re-absorbed or scattered away.
Thus, we should allow for multiple injection attempts. These
are accounted for by the term $b_\chi e^{i\omega T_\chi}$ in the
denominator.

As all the energy that leaves the cavity has to enter again in an
equilibrium situation, the total scattering amplitudes must obey
\begin{equation}\label{S=1}
  |S_\chi|^2=1, 
\end{equation}
which yields 
\begin{equation}
  |s_\chi|^2 + |a_\chi|^2+ 2 \mathrm{Re}\, s_\chi^* a_\chi e^{i\omega
    T_\chi} = 1 + |b_\chi|^2 + 2 \mathrm{Re}\, b_\chi e^{i\omega T_\chi}.
\end{equation}
Separating the slowly from the rapidly varying terms,
\begin{equation}\label{lowfast}
  |s_\chi|^2 + |a_\chi|^2 = 1 + |b_\chi|^2, \quad 
  \mathrm{Re}\, s_\chi^* a_\chi e^{i\omega T_\chi} = \mathrm{Re}\, b_\chi
  e^{i\omega T_\chi},
\end{equation}
we obtain
\begin{equation}\label{ab}
  a_\chi=e^{i\delta_\chi},\quad b_\chi = s_\chi^*e^{i\delta_\chi},
\end{equation}
where $\delta_\chi$ are a slowly varying phases. The normal modes of the
system in equilibrium are not given by Eq. (\ref{normal}) but by 
\begin{equation}\label{Normal}
 D=1-S_b S_c=0,
\end{equation}
which may be recast as
\begin{eqnarray}\label{phase}
  &&2\mathrm{arg}\left( 1+s_b e^{i\delta_b}e^{i\omega T_b}\right) +
  2\mathrm{arg}\left( 1+s_c e^{i\delta_c}e^{i\omega T_c}\right)
    \nonumber
  \\
  &&\quad
  + (\delta_b+\delta_c) + \omega(T_b + T_c)
   = 2 \pi n,
\end{eqnarray}
with  integer $n$. The first two terms in the LHS of
Eq. (\ref{phase}) oscillate rapidly, but are bounded within the
interval $(-\pi,\pi)$, while the second term is slowly varying. Thus,
the separation between nearby modes $\omega_n$ is close to  $\Delta
\omega^0=2\pi/(T_b+T_c)$ 
and the density of modes diverges in the limit $T_b,
T_c\to\infty$. This is to be expected, as our delayed re-radiation
accounts implicitly for the interaction with a thermal bath which has infinite
degrees of freedom. Our dissipative system together with the thermal
bath forms an extended closed system whose modes are actually real
\cite{sernelius06} and form a quasi-continuum. Our approach above is an
alternative to the introduction of ancillary systems to account for
dissipation \cite{barash75,njp}.

\begin{figure}
  \centering{\includegraphics{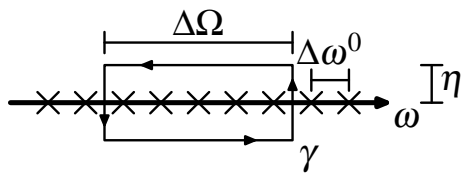}}
  \caption{\label{gamma} Integration contour $\gamma$ employed to obtain the
    density of states. We indicate the normal modes (crosses) separated
    approximately by $\Delta w^0$. }
\end{figure}

The actual number of modes $\Delta N$ within a small frequency range $\Delta
\Omega$ may be obtained using Cauchy's argument principle 
\begin{equation}\label{argument}
  \Delta N = \frac{1}{2\pi i} \oint_\gamma \frac{d}{d\omega} \log f(\omega),  
\end{equation}
where $\gamma$ is a contour that encircles counterclockwise the
interval $\Delta \Omega$, and $f(\omega)$ is an analytical function
that has the same zeroes as $D$ (Eq. (\ref{Normal})) and no poles
within $\gamma$. We choose a contour $\gamma$ that moves
towards the right a distance $\eta$ below the real
axis, then crosses the axis, moves back a distance $\eta$ above the
real axis, and finally crosses the axis to closes upon itself \cite{sernelius06}
(Fig. \ref{gamma}). 
Choosing as $f(\omega)$ the analytical continuation from the real axis
unto the complex plane of 
\begin{eqnarray}\label{fw}
  f(\omega)&=&(1+s_b e^{i\delta_b} e^{i\omega T_b}) (1+s_c e^{i\delta_c}
  e^{i\omega T_c})
  \nonumber
  \\
  &&
  - (s_b+ e^{i\delta_b} e^{i\omega T_b}) (s_c +
  e^{i\delta_c} e^{i\omega T_c}),
\end{eqnarray}
we obtain
\begin{eqnarray}\label{DeltaN}
  \Delta N &=& \left .\frac{1}{2\pi i} \Delta f(\omega) \right
  |_{\omega+i\eta}^{\omega-i\eta}
\nonumber
\\
&=& \frac{1}{2\pi i} \Delta \Bigl \{\log\left [-
    \frac{1 - s_b^* s_c^*}{1-s_b s_c}\right ] + i(\delta_b+\delta_c) 
\nonumber
\\
&& +  i\omega(T_b+T_c) + \eta(T_b+T_c)\Bigr \}
\nonumber
\end{eqnarray}
in the limit $T_\chi\to\infty$, $\eta\to0$, $\eta T_\chi\to\infty$.
Subtracting the number  $\Delta N_0$ of modes corresponding to
vacuum and the thermal  reservoir only, obtained from
Eq. (\ref{DeltaN}) by replacing $s_b\to0$, $s_c\to1$, we obtain
\begin{equation}\label{DNDN0}
 \Delta N - \Delta N_0 = -\frac{1}{\pi}\Delta \mathrm{Im}
 \log(1-s_b s_c) \equiv \rho(\omega) \Delta\Omega,
\end{equation}
where we identify the contribution of the scatterers to the density of
states,
\begin{equation}\label{rho}
  \rho(\omega) = -\frac{1}{\pi}\mathrm{Im} \frac{d}{d\omega}\log(1-s_b s_c),
\end{equation}
for each value of $l,m,\zeta$.

From the density of states, one can proceed to calculate all the
thermodynamic quantities. For example, multiplying $\rho(\omega)$ by
the ground state energy $\hbar\omega/2$ of an oscillator of frequency
$\omega$, integrating over all frequencies and adding over all angular
momenta and polarizations we obtain the contribution of the scatterers
to the ground state energy,
\begin{equation}\label{U0}
  U_0=\frac{\hbar}{2\pi} \sum_{l,m,\zeta} \int_0^\infty du\,
  \log\left|1-s_{b,lm}^\zeta(iu) s_{c,lm}^\zeta(iu) \right |,
\end{equation}
where we have already rotated the integration trajectory unto the
imaginary axis.

Eq. (\ref{U0}) and similar equations easily derived for other
thermodynamic quantities are the main results of the present paper. In
order to evaluate them the only requirement is knowledge of the
scattering amplitudes corresponding to the surface of the cavity and
to the scatterer at the center.

\section{Empty cavity within a uniform medium}

For an empty cavity $s_{c,lm}^\zeta=1$, as the incoming field
becomes an outgoing field after crossing the origin. If
the cavity is surrounded by a uniform medium with a dielectric
function $\epsilon(\omega)$, then $s_{b,lm}^\zeta$ may be easily found
by writing the field within the cavity as a linear combination of
outgoing and ingoing fields (Eq. (\ref{Flm})), writing the field
within the medium as an outgoing field and matching both solutions
through the usual boundary conditions, i.e., the continuity of the
projections of both the electric and magnetic field along the
surface. The result \cite{unpub} is simply given by
\begin{equation}\label{sTE}
  s_{b,lm}^{TE}(\omega)=-\frac{Q_l^{MoVo}-Q_l^{VoMo}}{Q_l^{MoVi}-Q_l^{ViMo}}
\end{equation}
where
\begin{equation}\label{Ql}
  Q_l^{AdA'd'}\equiv k_{A'}\, h_l^d(k_A R)\, \hat D
  h_l^{d'}(k_{A'}R),\ A,A'=V,M,\ d,d'=i,o,
\end{equation}
$k_A$ is the wavenumber within vacuum ($A=V$, $k_V=k=\omega/c$) or
within the medium 
($A=M$, $k_M=k \sqrt\epsilon$), $h_l^d$ are the ingoing ($d=i$) or
outgoing ($d=o$) spherical Hankel functions, and $\hat D$ is the operator
\begin{equation}\label{Dg}
  \hat D g(x) \equiv g'(x) + g/x.
\end{equation}
Similarly, for TM polarization we have
\begin{equation}\label{sTM}
  s_{b,lm}^{TM}(\omega)= - \frac {Q_l^{MoVo}-Q_l^{VoMo} / \epsilon}
  {Q_l^{MoVi} - Q_l^{ViMo}/\epsilon}.
\end{equation}
As $s_{b,lm}^\zeta$ is independent of $m$, we may replace the sum over
$m$ in Eq. (\ref{U0}) by a factor of $2l+1$.

\begin{figure}
  \centering{\includegraphics{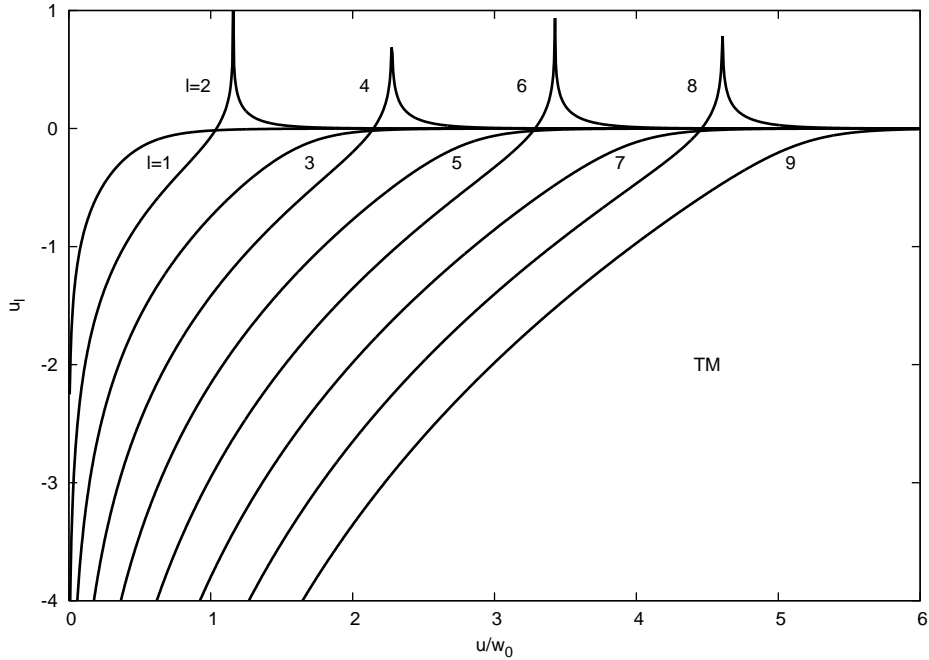}}
  \caption{\label{fsTM} TM contribution
    $u_l=(1/2\pi)\log|1-s_{b,lm}^{TM}|$ to the energy $U_0$ of a
    cavity of radius $R=c/\omega_0$ within a host with a Lorentzian
    dielectric response
    $\epsilon=1+\omega_p^2/(\omega_0^2-\omega^2-i\gamma\omega)$, with
    $\omega_p=\omega_0$ and $\gamma=0.01\omega_0$. 
    }
\end{figure}

In Fig. \ref{fsTM} we show the TM contribution to the integrand
of Eq. (\ref{U0}) as a function of the
imaginary part of the frequency $u=\omega/i$ calculated for a
dielectric cavity with a Lorentzian response
$\epsilon=1+\omega_p^2/(\omega_0^2-\omega^2-i\gamma\omega)$.
The figure shows integrable singularities at $u=0$. A second
singularity is seen for even values of $l$. Nevertheless, it is
clearly seen that the contributions to the energy grow with $l$ and
that Eq. (\ref{U0}) doesn't converge. 

Our calculation above includes a realistic dielectric response for the
medium surrounding the cavity and in our calculation the
electromagnetic field permeates all space. Thus, the singularities
above are of a different physical nature than the singularities arising in
naive calculations for flat surfaces. For example, the singularities
in the Casimir force among perfect conducting slabs may be removed by
introducing a high frequency cutoff which accounts for the high
frequency transparency of real metals and when the mechanical
properties of the field beyond the slabs is accounted for. Those
recipes wouldn't cure the present divergence.

Notice that large $l$'s correspond to spatial oscillations around the
sphere with a small lengthscale $d=2\pi R/l$. As we may
assume a  smallest lengthscale $d_{\mathrm{at}}$ of atomic
dimensions, it seems reasonable to impose a corresponding cutoff at
$l_{\mathrm{max}}=2\pi R/d_{\mathrm{at}}$. A careful analysis
\cite{unpub} shows that in this case Eq. (\ref{U0}) converges, but
that the leading terms are of order $(R/d_{\mathrm{at}})^3$ and
$(R/d_{\mathrm{at}})^2$, i.e., of the order of the number of atoms
within the volume and the surface of a sphere of radius
$R$. Mathematically, these terms arise from the leading terms in an
expansion of the integrand of Eq. (\ref{U0}) for large $l$, which,
after summing over $m$ are of order $l^2$ and $l$ (with logarithmic
corrections). Physically, the 
reason is that Eq. (\ref{U0}) includes the electromagnetic interaction
between nearest neighbor atoms, which in a continuum description are
infinitely small and infinitely close to each other. Similar divergent
terms were found in a pairwise perturbative calculation \cite{barton01}
for cavities surrounded by a very diluted dielectric. As argued in
Ref. \cite{barton01}, these short range contributions should be taken
into account before conclusions about the sign of the Casimir force
can be drawn.

A finite Casimir energy may be obtained by subtracting from our
result above those terms that contribute to the divergence before we
take the limit  $d_{\mathrm{at}}\to0$. Nevertheless, the
resulting energy would be an experimentally inaccessible
quantity for a spherical cavity, as 
changing the cavity radius would require adding or removing
atoms or else introducing strains that would
produce an elastic stress \cite{hertzberg05}. Furthermore, a full
calculation of the 
interatomic interactions would be required in order to compare
experimental results to theoretical calculations, and our
Eq. (\ref{U0}) would be insufficient. One way out of these
difficulties is to study other geometries where motion introduces no
elastic stresses. For example, it has been found \cite{hertzberg05} that for a
piston sliding along a cylinder, the contribution of its position to the
Casimir energy is free of cutoff dependent singularities and is
attractive.

\section{Atom within a cavity}

On the other hand, Eq. (\ref{U0}) would still be useful in situations where
the geometry of the cavity is left unchanged. For example, it may be
used to calculate the change 
\begin{equation}\label{DeltaU0}
  \Delta U=\frac{\hbar}{2\pi} \sum_{l,m,\zeta} \int_0^\infty du\,
  \log\left|\frac{1-s_{b,lm}^\zeta(iu) s_{c,lm}^\zeta(iu)}
		 {1-s_{b,lm}^\zeta(iu)} \right |,
\end{equation}
in the energy of the system when an atom is introduced at the center of the
cavity. Using the usual selection rules, we obtain that the dispersion
amplitude for an atom, $s_{c,lm}^\zeta=1$, would be the same as for empty
space, unless $l=1$ and the polarization $\zeta=\mathrm{TM}$, in which
case,
\begin{equation}\label{salpha}
  s_{c,1m}^{TM}= \frac {1+\frac{2}{3}i k^3 \alpha}{1-\frac{2}{3}i k^3 \alpha},
\end{equation}
where $\alpha$ is the electric-dipole polarizability of the atom. Therefore,
there is only one finite term in Eq. (\ref{DeltaU0}).

\begin{figure}
  \centering{\includegraphics{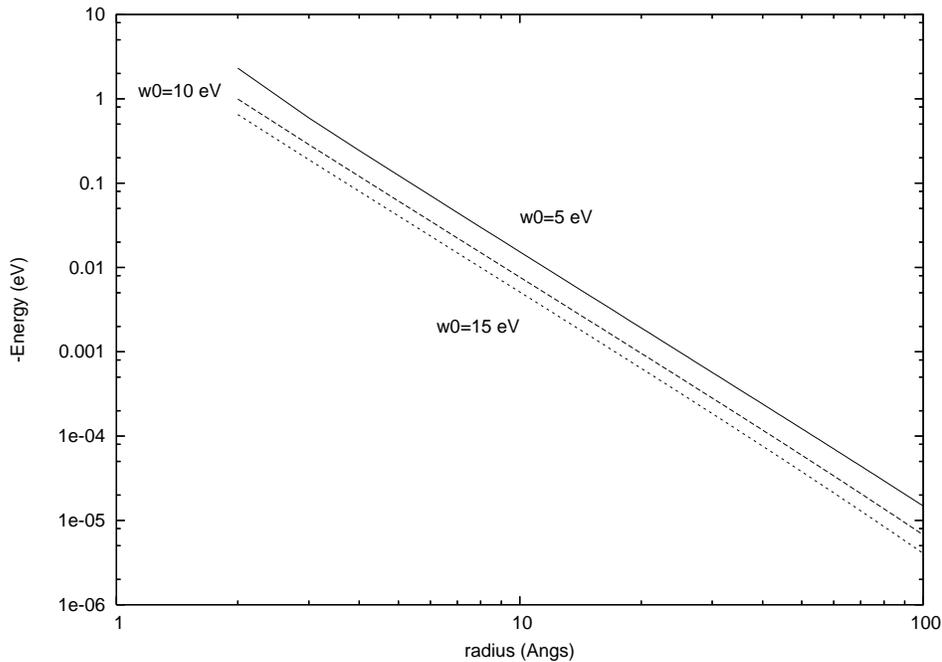}}
  \caption{\label{figDU}Contribution to the energy of a polarizable
    atom lying at the center of a perfectly conducting cavity as a function
    of the radius $R$ of the cavity. The polarizability is taken as a
    Lorentzian with a resonance frequency $\omega_0$. 
}
  
\end{figure}

In Fig. \ref{figDU} we show the contribution to the energy of a cavity
from an atom lying at its center. The atomic polarizability is taken
as a Lorentzian
\begin{equation}\label{alpha}
  \alpha(\omega)=\frac{e^2/m}{\omega_0^2-\omega^2}
\end{equation}
with a single resonance frequency $\omega_0$. The energy is negative
and proportional to $R^{-3}$ and is of the order of tens of
meV's for radii of a few nanometers.

\section{Conclusions}
Based on a scattering approach we have derived an expression for the Casimir energy of a 
spherical cavity with dispersive and absorptive dielectric properties. It turns out that
the expression is divergent due to the summation over angular momenta, independently of
the dielectric behavior of the cavity walls. This may be regularized 
by imposing a cutoff associated with the finite separation of atomic scatterers forming the
boundary of the cavity, leading to contributions proportional to the number of atoms in the volume
and surface of the cavity. On the other hand, the energy shift for an atomic radiator placed
in the center of the cavity is finite, since it does not require to perform virtual work to
modify the geometry of the cavity

\section*{Acknowledgments}
We acknowledge useful discussions with Jos\'e R\'ecamier and with
Jos\'e C. Torres-Guzm\'an. This work was partially funded by DGAPA-UNAM
under grants No. IN111306 and IN118605.

\end{document}